\documentclass[aps,twocolumn,amsmath,amssymb]{revtex4}
\usepackage{graphicx}
\usepackage{epsfig}
\usepackage{bm}
\usepackage{amsmath}
\usepackage{amssymb}

\def\LB{\left(}
\def\RB{\right)}
\def\ba{\begin{eqnarray}}
\def\ea{\end{eqnarray}}

\def\ua{\uparrow}
\def\da{\downarrow}

\newcommand{\be}{\begin{equation}}
\newcommand{\ee}{\end{equation}}
\newcommand{\bea}{\begin{eqnarray}}
\newcommand{\eea}{\end{eqnarray}}

\begin{document}
%\mbox{}
%\vspace{-2.0cm}
%     {\flushright{\it Draft. Do not distribute. }}
%\vspace{2.0cm}

\title{Quantum Electrical Dipole in Triangular Systems: \\ a Model for
Spontaneous Polarity in Metal Clusters}

\author{Philip B. Allen$^{1,2}$}
\author{Alexander G. Abanov$^{1}$}
\author{Ryan Requist$^{1}$}

\affiliation{$^{1}$Department of Physics and Astronomy, Stony Brook University,
\\
Stony Brook, NY 11794-3800, U.S.A.\\
$^{2}$Department of Applied Physics and Applied Mathematics\\
Columbia University, New York, NY 10032, U.S.A.}

\date{\today}

\begin{abstract}
Triangular symmetric molecules with mirror symmetry
perpendicular to the 3-fold axis are forbidden 
to have a fixed electrical dipole moment.  However,
if the ground state is orbitally degenerate and lacks
inversion symmetry, then a ``quantum''
dipole moment does exist.  The system of 3 electrons
in D$_{3h}$ symmetry is our example.  This system is realized
in triatomic molecules like Na$_3$.  Unlike the
fixed dipole of a molecule like water, the quantum moment does not
point in a fixed direction, but lies in the plane
of the molecule and takes quantized values $\pm \mu_0$
along any direction of measurement in the plane.
An electric field $\vec{F}$ in the plane leads to a linear
Stark splitting $\pm\mu_0|\vec{F}|$. We introduce a toy model to study the
effect of Jahn-Teller distortions on the quantum dipole moment. We find that
the quantum dipole property survives when the dynamic Jahn-Teller effect is
included, if the distortion of the molecule is small.  Linear
Stark splittings are suppressed in low fields by molecular
rotation, just as the linear Stark shift of water is
suppressed, but will be revealed in moderately large
applied fields and low temperatures.
Coulomb correlations also give a partial suppression.
\end{abstract}

\maketitle

%%%%%%%%%%%%%%%%%%%%%%%%%%%%%%%%%%%%%%%%%%%%%%
%%%%%%%%%%%%%%%%%%%%%%%%%%%%%%%%%%%%%%%%%%%%%%
\section{Introduction}\label{sec:intro}

Moro {\it et al.} (de Heer's group \cite{Moro1,Moro2})
found an unexpected low temperature
electrical dipole, apparently not classical,
in nearly all clusters of $N$ niobium atoms with
$2<N<150$.  Here we use the X$_3$ molecule 
to propose an answer to the question, what a non-classical 
electrical dipole might be.  Our model
electrical dipole bears a
close analogy to the non-classical
magnetic moment of a spin 1/2 particle.
It is meaningless to ask 
the direction of the moment of an isolated 
electron, because different components of the moment operator
do not commute.  In a field $\vec{B}$ 
the degeneracy with respect to spin degrees of freedom 
is lifted and the ground state has energy
$E_0=-\mu_B |\vec{B}|$ and is characterized
by having maximal projection of the
moment $\vec{\mu}_M=\langle 0|\hat{\vec{\mu}}_M|0\rangle
=\mu_B\hat{B}$ along the 
magnetic field.

If the electron is not isolated but in
contact with a heat bath, the statistical
average of the moment is $\vec{\mu}_M=
\mu_B \tanh(\mu_B |\vec{B}|/k_B T)\hat{B}$.
The susceptibility has the Curie form
$\chi=\mu_B^2/k_B T$.
The thermal behavior at weak applied fields
is classical \cite{Vanvleck}, with a well-defined linear response 
$\vec{\mu}_M=\chi\vec{B}$.  The operational
definition of a dipole is
a Curie response of the isolated system, and the moment is found
from $\sqrt{\chi T}$.  Classically we
know that the moment points at some angle
$\theta$ to the field and precesses.
Quantum mechanics forces a more abstract view,
which replaces the direction of the moment 
by the ``quantization axis.''

Usually we think of an electrical dipole moment $\vec{\mu}$
on a molecule in classical terms.  When rigidly
fixed in space, the moment has a value which
is not quantized in atomic units $ea_B$.  Its
value is an accidental result of chemistry.
In the frame of the molecule,
the components of the electric dipole moment operator commute,
so a direction can be defined in this frame.
A thermalized vapor of molecules has moments randomly
oriented.  In an electric
field $\vec{F}$ there is a net polarization proportional
to $\vec{F}$, corresponding
to an average dipole per molecule $\chi\vec{F}$,
with the Curie form $\chi=\mu^2/3k_B T$ \cite{Vanvleck}.
Quantum mechanics enters in the usual picture when
$k_B T$ is small enough that the spacing of quantized rotational
levels $E_{\rm rot}=(\hbar^2/2I)L(L+1)$ is
no longer negligible.  The ground state $L=0$
has full rotational symmetry with zero average
dipole, and the Curie divergence is cut off,
replaced by $\chi\rightarrow\mu^2/(\hbar^2/2I)$\cite{Vanvleck,Townes}.

More interesting quantum effects, including
non-commutation of the relevant matrix elements
of the electric dipole operator, may occur
when the molecule has an orbitally degenerate
ground state, and lacks an inversion center.
We consider the simplest case, with equilateral
triangular (D$_{3h}$) symmetry.
A {\em fixed} (classical) electrical dipole
is inconsistent with D$_{3h}$ symmetry.
To quote Townes and Schawlow, ``there can be no
dipole moment perpendicular to the axis of a symmetric top.''
\cite{Townes1}.  
This statement is also true in quantum mechanics
if the ground state is non-degenerate.
However, there is a non-zero electrical
dipole matrix in the doubly degenerate $E$ representation
of D$_{3h}$.  This matrix has non-commuting
Cartesian components.  The moment exhibits quantum behavior
similar to the spin of an electron.
Further quantum effects suppress this moment,
similar to the way zero-point rotation suppresses
the usual molecular moment \cite{Vanvleck,Townes,Lehmann},
but there are very interesting differences.
Orbital magnetic moments also occur in this model, and
are discussed in Appendix \ref{app:bloch}.

Our model (Sec. \ref{sec:trimer}) is a simplified picture of
an alkali trimer like Na$_3$.
We start with a rigid equilateral triangle
geometry, fixed in space,
representing, for example, three atoms.  All electrons
except the outer 3 valence electrons (one
per atom) are frozen.  One alternate realization is 
the ``R-center'' in an alkali halide crystal
\cite{Stoneham}.  In crystalline NaF, for example, the center has
three adjacent F$^-$ vacancies with equilateral
triangular symmetry in a (111) plane.  Each vacancy
binds one electron donated by neighboring Na$^+$
ions.  Another alternate realization is a quantum
well with triangular shape \cite{Zarand}, gated to hold
three electrons, or three identical closely-coupled
quantum dots in triangular geometry, gated
to hold one electron each.

Because we want our model to be relevant to
Nb$_N$ clusters, we must add some additional physics.  We consider 
all distortions of the triangle in linear
approximation (Sec. \ref{sec:adiabatic}).  These have a
particular importance because of a Jahn-Teller
instability and a Longuet-Higgins-Berry phase
that inverts the order of singly and doubly
degenerate levels (Sec. \ref{sec:vibrations}), and restores
the ground-state degeneracy.  We also consider rotations
in space, but, for simplicity, only in two
dimensions (Sec. \ref{sec:rotations}).  Additional complications
coming from higher order couplings, from
rotations in the third dimension (these do
not alter anything fundamental), and from
antisymmetry of the nuclear wavefunction
(the Na nucleus is a spin 3/2 Fermion)
are omitted for now.  We plan to  discuss some of these 
in a subsequent paper \cite{xxx}
which will focus on the Na$_3$ trimer.
  
%%%%%%%%%%%%%%%%%%%%%%%%%%%%%%%%%%%%%%%%%%%%%%
%%%%%%%%%%%%%%%%%%%%%%%%%%%%%%%%%%%%%%%%%%%%%%
\section{Dipole moment of undistorted trimer}\label{sec:trimer}

The minimal toy model has three electrons which live
in a Hilbert space of three $s$ orbitals $|n\rangle, \ n=0,1,2$
located on ``atoms'' which define vertices of
a triangle.  This crudely represents the outer electrons of
an alkali or noble metal
trimer X$_3$.  The subsequent paper will make contact with
the specific case of Na$_3$.

\begin{figure}[t]
   \includegraphics[width=\columnwidth]{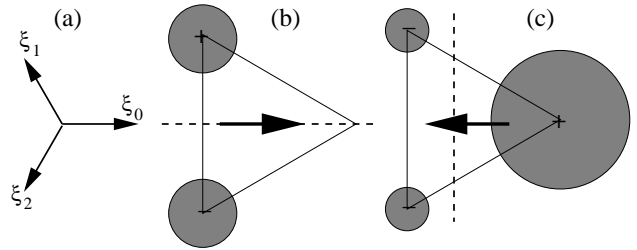}
   \caption{\label{fig:dipo} 
(a) The unit vectors $\hat{\xi}_n$ which locate the undistorted
positions in the molecular frame.  (b) The conventional real 
basis function $\psi_2$ and (c) $\psi_3$ for the doubly degenerate
$E^{\prime}$ orbitals which diagonalize $H_{\rm el}$ when the
molecule is undistorted.
Nodal planes are shown as dashed lines.  If either of these states
is occupied once and the other is empty, the atom will have
a dipole moment $\pm \mu_0$, where $\mu_0=er_0 /2$ and
$r_0$ is the radial distance from the center to any atom.  This
dipole is shown as the bold vector.}
\end{figure}

We define positions
of X nuclei as $\vec{R}_n,  \ n=0,1,2$.  We write $\vec{R}_n
= \hat{\mathbf{\Omega}}\vec{r}_n$, where
$\vec{r}_n = \vec{r}_{n0}+\vec{u}_n$ 
give positions of nuclei in an internal reference frame 
of the molecule.  The orientation of the molecular frame with
respect to the laboratory frame is given by an $SO(3)$ matrix
$\hat{\mathbf{\Omega}}$ which can be parameterized by three Euler angles.
We choose the internal coordinates
$x,y,z$ such that the molecule lies in the $xy$ plane.
The atoms have small displacements $\vec{u}_n$
relative to equilateral triangular positions
$\vec{r}_{n0}=r_0 \hat{\xi}_n$.  The equilateral vertices
are labeled by the unit vectors $\hat{\xi}_n$ where
$\hat{\xi}_0$ lies in the molecular $\hat{x}$ direction, and 1,2
are rotated by $\pm 120^{\circ}$ around $\hat{z}$ 
as shown in Fig. \ref{fig:dipo}.

The electrons are described by the tight-binding model
\be
    H_{\rm el} = -\sum_{n=0}^2 (t_{n,n+1} c_n^\dagger c_{n+1} +c.c.) +H_{\rm int},
 \label{eq:hel}
\ee
where $c_n$ is the annihilation operator of an electron on the 
$s$-orbital of the $n$-th atom.  Coulomb interactions are mostly
ignored.  The effects of an on-site Hubbard interaction 
$H_{\rm int}$ are considered in Appendix \ref{app:hubbard}.
The hopping matrix elements $t_{n,n+1}$ depend on the
distance between atoms $|\vec{r}_n -\vec{r}_{n+1}|$.

The interaction with the external electric field
$\vec{F}$ can be represented by
\be
    H_F = -\vec{F}\cdot(\hat{\mathbf{\Omega}}\vec{\mu})
= -(\hat{\mathbf{\Omega}}^{-1}\vec{F})\cdot \vec{\mu}
=-\vec{F}_m \cdot \vec{\mu},
 \label{eq:hf}
\ee
where the electric dipole moment in the molecule reference frame is given by
\be
    \vec{\mu} = -\sum_{n=0}^2 e\vec{r}_n c_n^\dagger c_n.
 \label{eq:dop}
\ee
The factor $\vec{F}_m=\mathbf{\hat{\Omega}}^{-1}\vec{F}$ is the electric field
$\vec{F}$ as seen in the molecule reference frame.

For the rest of this section we take the molecule to be
stationary, so that the lab frame coincides with the molecule
frame ($\hat{\mathbf{\Omega}}=1$), and neglect distortions
($\vec{u}_n=0$).  Therefore $t_{n,n+1}$ is a constant, $t_0$.
For Na$_3$, the hopping matrix element $t_0$ is about 0.4 eV
\cite{Cocchini}.  Eq. (\ref{eq:hel}) is the $N=3$ case of the 
symmetric $N$-atom ring, treated in H\"uckel theory.
The states follow from $N$-fold rotational symmetry
\begin{equation}
    |k\rangle=\frac{1}{\sqrt{N}}\sum_n e^{i2\pi kn/N}|n\rangle.
 \label{eq:bloch}
\end{equation}
The dimensionless wavenumber $k=-1,0,1$, defined {\it modulo} $N=3$,
labels the energy levels.  The energy eigenvalues are
$\epsilon_k=-2t_0\cos(2\pi k/N)$.  The lowest state $k=0$, of energy
$-2t_0$, is bonding (has no nodes), and is twice
occupied.  The other two states, $k=\pm 1$, are degenerate
with energy $+t_0$ and hold one electron.
                                                                                
Now consider the electrical dipole operator (\ref{eq:dop}).
For a single electron in the orbital $|n\rangle$,
the dipole operator has expectation value 
$\langle n|\vec{\mu}|n\rangle =-e\vec{r}_{n0}$.  The eigenstates
$|k\rangle$ have equal charge on all atoms and therefore zero net dipole.
However, there are off-diagonal matrix elements obeying a
selection rule $\Delta k=\pm 1$,
\begin{equation}
    \langle k+1|\vec{\mu}|k\rangle=-\mu_0 (\hat{x}-i\hat{y}),
 \label{eq:mu}
\end{equation}
where $\mu_0 = er_0 /2$, and $\hat{x},\hat{y}$ are unit vectors 
in the molecule reference frame.
The matrix element $\langle k|\vec{\mu}|k+1\rangle $
is the complex conjugate of the one shown.
A special property of the trimers with half-filled valence shells
is that there are part-filled degenerate states with $k=\pm 1$.  The
selection rule $\Delta k=\pm 1$ does not exclude the matrix element
$\langle k=-1|\vec{\mu}|k=1\rangle $ because -1 is the same as 2.  Thus
there is an off-diagonal dipole
matrix element in the degenerate electronic ground state manifold.
The $-\vec{\mu}\cdot\vec{F}_m$ coupling term, Eq. (\ref{eq:hf}), is
a $2 \times 2$ matrix in the subspace $|k=-1\rangle,|k=1\rangle$, namely
\begin{equation}
    {\cal H}_{F}^{\rm undist}
    =-\mu_0 \vec{\sigma}\cdot\vec{F}_m
 \label{eq:mudotf}
\end{equation}
where $\vec{\sigma}=\sigma_x \hat{x}+\sigma_y \hat{y}$ and
$\sigma_x$ and $\sigma_y$ are the usual off-diagonal Pauli
matrices.  The eigenvalues split linearly, $\Delta\epsilon=\pm\mu_0
|\vec{F}|$, where it is important that the component $F_{mz}$ of $\vec{F}$
perpendicular to the plane of the molecule is omitted (that is,
$|\vec{F}|$ is $(F_{mx}^2+F_{my}^2)^{1/2}$.)

One might expect that other ring molecules with odd numbers
of atoms, like C$_5$H$_5$, would also have this effect.  Indeed,
there is a dipole in the degenerate manifold $|k=2\rangle,|k=-2\rangle$ for
this case (since -2 is the same as 3 {\it modulo} 5), but this
level is empty and there is no dipole in the partly filled
$|k=1\rangle,|k=-1\rangle$ manifold.  One might also be surprised that
the Cartesian elements of the electrical dipole operator are represented by
non-commuting $2\times 2$ matrices, since these are
commuting operators in a full theory.
In the full three-dimensional space of the three $s$ states, the matrices
$\mu_x,\mu_y,\mu_z$
commute, but in the $k=\pm 1$ degenerate subspace relevant to
low energy physics, they do not.
                                                                                
One can visualize this dipole moment by using
the real linear combinations $\psi_2=(|k=1\rangle-|k=-1\rangle)/\sqrt{2}i$
and
$\psi_3=(|k=1\rangle+|k=-1\rangle)/\sqrt{2}$,  the  conventional
$E^{\prime}$ basis
states with amplitudes $(0,1,-1)/\sqrt{2}$ and $(2,-1,-1)/\sqrt{6}$ on
atoms $(0,1,2)$ shown on Fig. \ref{fig:dipo}.  These states have diagonal
dipole expectation values $\pm\mu_0 \hat{x}$ and off-diagonal
elements $\mu_0 \hat{y}$.

%
%\begin{equation}
%{\cal H}_B^{\rm undist}=-\mu_{\rm orb}\sigma_z B_z
%\label{eq:muorb}
%\end{equation}
%
It is worth commenting on the magnetic properties of this model.
The unpaired electron has a net spin.  This couples weakly, through
the spin-orbit interaction, to orbital angular momentum.  The undistorted
molecule has a large orbital magnetic moment.  If the discrete three-membered
ring were replaced by a 1-d continuum (in the spirit of the free-electron
approximation for bulk Na metal), the orbital moment would be
$\vec{\mu}_{\rm orb}=k\mu_B \hat{z}$ where $k=\pm 1$ is the quantum
number of the outer unpaired electron.  The free electron approximation
gives the right symmetry and order of magnitude, but is not 
quantitatively correct.
A correct calculation will give $\vec{\mu}_{\rm orb}=k\gamma\mu_B \hat{z}$
where we estimate $\gamma\approx 0.3$ since the moment must be
approximately $2\pi e\omega A$ with $A$ the area of the molecule, 
and $2\pi\omega$
is the classical frequency of electron rotation, approximately 
$\omega=t_0/\sqrt{3}\hbar$.
In the next sections we find that distortions 
quench the orbital magnetic moment.
For this reason we are not concerned with computing $\gamma$ which would have
required additions to the model.  Unlike the magnetic dipole moment, distortions
reduce the electrical dipole moment only by a factor of 2.
The ``Bloch-sphere'' gives a convenient representation of the magnetic 
and electric dipoles in this model, and is
discussed in Appendix \ref{app:bloch}.

%%%%%%%%%%%%%%%%%%%%%%%%%%%%%%%%%%%%%%%%%%%%%%
%%%%%%%%%%%%%%%%%%%%%%%%%%%%%%%%%%%%%%%%%%%%%%

\section{Static distortions and Adiabatic Longuet-Higgins
phase}\label{sec:adiabatic}

To describe small distortions, consider the first order expansion
in $\vec{u}_n=\vec{r}_n-r_0 \hat{\xi}_n$.  The hopping matrix is
expanded to first order around the undistorted atom separation $\sqrt{3}r_0$,
\bea
    t_{n,n+1} &=& t_0-\frac{g}{\sqrt{3}}
    (|\vec{r}_n -\vec{r}_{n+1}|-\sqrt{3}r_0) \nonumber \\
    &=& t_0-\frac{g}{3}(\vec{u}_{n+1}-\vec{u}_n)\cdot
    (\hat{\xi}_{n+1}-\hat{\xi}_n),
 \label{eq:tex}
\eea
where $g$ is the electron-phonon coupling.

To completely specify three atomic coordinates $\vec{R}_n$ requires nine numbers.
Three give the position of the center of mass, and describe rigid
translations of the molecule which separate out and do
not contribute to Eq. (\ref{eq:tex}).
Three (e.g., Euler angles) locate
the spatial orientation of the plane of the molecule, and the
rotation angle of the molecule in the plane.  These 
rotations, which also do not contribute to Eq. (\ref{eq:tex}),
 will be discussed later.   Finally, three parameters,
chosen as amplitudes of three orthogonal normal modes of distortion,
give the vibrations which do appear in Eq. (\ref{eq:tex}).

\begin{figure}[t]
   \includegraphics[height=3.0in,width=3.0in,angle=0]{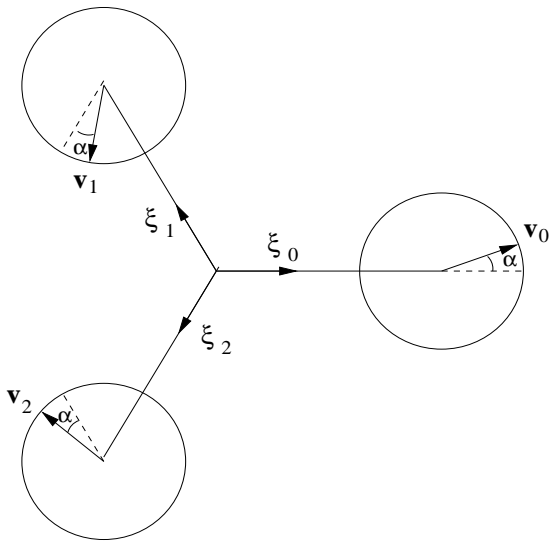}
   \caption{\label{fig:geom} Geometry of the distorted trimer.
Atoms are at coordinates $\vec{r}_n=(r_0 + Q_1)\hat{\xi}_n + Q \hat{v}_n$.
When $Q=0$ the molecule has full D$_{3h}$ symmetry.
When the angle $\alpha=0$, the vectors
$\vec{v}_n$ lie along the dashed lines and describe a symmetric
stretch which makes an isosceles distortion.  This is the mode
called Q$_2$ by Van Vleck.  When $\alpha=\pi/2$, the vectors
$\vec{v}_n$ give the asymmetric stretch Q$_3$.
These two patterns form the conventional basis for the
doublet $E^{\prime}$ vibrations.  Rather than $(Q_2,Q_3)$, we use the
parameters $(Q,\alpha)$ to describe the $E^{\prime}$ vibrations.  Note
that the three vectors $\vec{v}_n$ lie at 120$^{\circ}$ and add to
zero, independent of $\alpha$.  Note also that when $\alpha=120^{\circ}$,
the distortion is isosceles with atom 2 at the vertex, and when
 $\alpha=240^{\circ}$, the isosceles distortion has atom 1
at the vertex.}
\end{figure}

The normal modes of an equilateral triangle
comprise a symmetric ($A_1^{\prime}$, or ``breathing'') mode
and a doubly degenerate ($E^{\prime}$) mode.  Their amplitudes are
conventionally \cite{Vleck,Mead} called ($Q_1,Q_2,Q_3$).  Fig. \ref{fig:geom}
introduces a second trio of unit vectors $\hat{v}_n$.
Like the previously introduced set $\hat{\xi}_n$, these vectors
lie at 120$^{\circ}$ to each other; they add to zero and are orthogonal
to all translations and rotations and to the breathing mode eigenvectors
$\hat{\xi}_n$.

The atomic coordinates
omitting rotations and translations are
\bea
    \vec{r}_n &=& (r_0 +Q_1) \hat{\xi}_n
    +Q\hat{v}_n
 \label{eq:coord} \\
    \hat{\xi}_n &=&
    (\cos(2\pi n/3), \sin(2\pi n/3),0)^{\dagger}
 \label{eq:xi} \\
    \hat{v}_n &=&(\cos(\alpha-2\pi n/3),
    \sin(\alpha-2\pi n/3),0)^{\dagger}
 \label{eq:v}
\eea
where the dagger indicates the transpose of the row vector, which
is a column vector.  The amplitude $Q=\sqrt{Q_2^2 +Q_3^2}$ \cite{Note0} 
and angle $\alpha=\tan^{-1}(Q_3/Q_2)$, shown on Fig \ref{fig:geom},
parameterize the $E^{\prime}$ distortion. 
We will also assume that $Q_1,Q\ll r_0$ so that molecule
is close to its symmetric ``equilibrium'' (see below) configuration
$Q_1=Q=0$ or $\vec{r}_{n0} =r_0 \hat{\xi}_n$.

Vibrations are especially important because the orbitally degenerate ground
state is Jahn-Teller unstable.  The doublet $E^{\prime}$
distortions lift the degeneracy, lowering the energy of one of the doublet
$E^{\prime}$ electron states.  The latter is singly 
occupied in the ground state, thus lowering the energy of the molecule.
In this section we make the static approximation.  We assume that the
trimer is distorted so that $Q_{1,2,3}\neq 0$ and solve electronic part
of the problem in the presence of this static distortion.
First, project Hamiltonian (\ref{eq:hel}) onto the doubly degenerate ground
state of an undistorted trimer.  This projected Hamiltonian gives the
electron-phonon term acting in the subspace  of $(|k=-1\rangle,|k=+1\rangle)$
states as
\begin{equation}
    {\cal H}_{\rm ep}=gQ_1 {\bf {\sf 1}} - g(Q_2\sigma_x-Q_3\sigma_y).
 \label{eq:hep}
\end{equation}
The ``breathing'' amplitude $Q_1$ shifts the degenerate levels uniformly.
It also shifts the doubly occupied $|k=0\rangle$ level.  Both these shifts
will be absorbed into redefining the size
$r_0$ to be the true radius of the molecule.
The $(Q_2,Q_3)$ vibrations split the degenerate single-particle energies
by $\pm gQ$ where $Q$ is the magnitude
of the $(Q_2,Q_3)$ distortion.  The splitting does not depend
on the angle $\alpha$.  The ``lower Jahn-Teller''
electron orbital (which is the lower eigenstate of Eq.(\ref{eq:hep}),
occupied with unit probability
in adiabatic approximation) has the wavefunction
\begin{eqnarray}
    \psi_{\rm ad} &=& \frac{1}{\sqrt{2}}\left(
    e^{-i\alpha/2}|k=1\rangle+e^{i\alpha/2}|k=-1\rangle\right)
 \nonumber \\
    &=& \sin(\alpha/2)\psi_2+\cos(\alpha/2)\psi_3.
 \label{eq:psiad}
\end{eqnarray}
Here the first form of $\psi_{\rm ad}$ uses the basis of Eq.(\ref{eq:bloch})
while the second form uses the basis of real functions shown in
Fig. \ref{fig:dipo}.  For $\alpha=0$ the distortion is an acute
isosceles triangle and the lower Jahn-Teller state is $\psi_3$ as
shown in Fig. \ref{fig:dipo}, panel (c).  For $\alpha=180^{\circ}$,
the distortion is obtuse, and the lower state is $\psi_2$.  The
node of the lower state cuts the longer bond, leaving the shorter
bond nodeless.

The effective interaction with the external electric field can be obtained
from (\ref{eq:mudotf},\ref{eq:psiad}) as
\bea
    {\cal H}_F &=& \langle \psi_{\rm ad}|
    {\cal H}_{F}^{\rm rigid \ gs}
    |\psi_{\rm ad}\rangle
 \nonumber \\
    &=& -\mu_0 \vec{F}_m\cdot(-\vec{x}\cos\alpha
    +\vec{y}\sin\alpha).
 \label{eq:hefF}
\eea
The state $\psi_{\rm ad}$ has an electrical dipole
$\langle\psi_{\rm ad}|\vec{\mu}|\psi_{\rm ad}\rangle=\mu_0 \hat{P}$, where
the unit vector $\hat{P}$ is $(-Q_2 \hat{x}+Q_3 \hat{y})/Q$.
A correction of order $(Q/r_0)\cos(3\alpha)$ to the dipole
will also occur because the values of the vectors $\vec{r}_n$ in 
Eq. (\ref{eq:dop}) change under distortion.  However, effects ignored
in our model make similar size corrections, so we ignore this.
The dipole magnitude $\mu_0$ is the same as for the undistorted D$_{3h}$ case,
but it points in a fixed direction.  For $\alpha=0$, the distortion
is acute and the ground state wavefunction the same as Fig. \ref{fig:dipo}(c),
with $\vec{\mu}$ in the $-\hat{x}$ direction.  As $\alpha$
increases in a counter-clockwise sense, the negative charge of the
last electron circulates clockwise (corresponding to current moving
counterclockwise).  The dipole $\vec{\mu}$ counter-rotates as
$\alpha$ rotates.  We have lost the quantum nature of
the dipole, which now orients classically.  However,
quantizing the distortional degrees of freedom will restore the quantum aspect. 
The orbital magnetic dipole is quenched in the adiabatic ground state,
since the $k=\pm 1$ components are included in $\psi_{\rm ad}$ 
with equal magnitude.
Quantizing distortions will restore the magnetic dipole only very weakly.

Notice that the adiabatic wavefunction Eq. (\ref{eq:psiad}) changes
sign when the molecule is taken once around the point $Q=0$ of
``conical intersection'' of the upper and lower Jahn-Teller energy surfaces.
That is, when $\alpha$ increases by $2\pi$, $\psi_{\rm ad}$ changes sign.
This adiabatic phase $\pi$ picked up by $\psi_{\rm ad}$ is known as
the Longuet-Higgins phase\cite{Herzberg} or as Berry's
phase\cite{Shapere}.
Important consequences of this for the quantized distortional states
and for the electric dipole are worked out in the next section.

%%%%%%%%%%%%%%%%%%%%%%%%%%%%%%%%%%%%%%%%%%%%%%
%%%%%%%%%%%%%%%%%%%%%%%%%%%%%%%%%%%%%%%%%%%%%%
                                                                                         
\section{Quantized vibrations altered by Longuet-Higgins
phase}\label{sec:vibrations}

The full Hamiltonian of the molecule can be written as
\be
    {\cal H} = H_{\rm ions} +H_{\rm pot} +H_{\rm el} +H_F,
 \label{eq:hfull}
\ee
where
\be
    H_{\rm ions} = -\frac{\hbar^2}{2M}
    \sum_{n=0}^{2}\frac{\partial^2}{\partial \vec{R}_n^2},
 \label{eq:hions}
\ee
is the kinetic energy of the Na$^+$ ions.  The potential energy
is approximated by central springs,
\bea
    H_{\rm pot} &=& \sum_{n=0}^2 \frac{K}{2}(|\vec{r}_n-\vec{r}_{n+1}|-\sqrt{3}r_0)^2
\nonumber \\
&\approx& \frac{K}{2}\left[ 3Q_1^2 + \frac{9}{2}Q^2 \right].
 \label{eq:hpot}
\eea
We use cyclic notations $n+3\equiv n$;
$H_{el}$, $H_F$ are Hamiltonians (\ref{eq:hel},\ref{eq:hf}) of the valence 
electrons and of coupling to an external electric field $\vec{F}$.

In adiabatic approximation one looks for the total wave function of
electrons
and ions in the form of a product $\Psi(\vec{R}_n)\psi_{\rm ad}$.
Then the effective Hamiltonian acting on $\Psi(\vec{R}_n)$ is obtained
from Eqs. (\ref{eq:hep},\ref{eq:hions},\ref{eq:hpot}). It acquires an induced
``potential energy'' $\langle \psi_{\rm ad}|{\cal H}_{\rm ep}|\psi_{\rm
ad}\rangle = gQ_1-gQ$
and can be written in the reference frame of the molecule
as the sum of the kinetic and potential energy of vibrational modes,
\begin{eqnarray}
    {\cal H}_{\rm ad}&=& {\cal H}_{A}+ {\cal H}_{E^{\prime}} \nonumber \\
    {\cal H}_{A}&=& \frac{3M}{2}\dot{Q}_1^2 + \frac{3K}{2}Q_1^2 +gQ_1 \nonumber \\
          &\rightarrow&\frac{3M}{2}\dot{Q}_1^2 + \frac{3K}{2}Q_1^2 -\frac{g^2}{6K}
      \label{eq:ha} \\
    {\cal H}_{E^{\prime}}&=& \frac{3M}{2}(\dot{Q}_2^2+\dot{Q}_3^2)
     +\frac{9K}{4}Q^2  - gQ,
 \label{eq:he}
\end{eqnarray}
where $M$ is the mass of the Na atom.  The second form 
of ${\cal H}_A$, Eq. (\ref{eq:ha}), results from the 
redefinition $r_0-g/3K \rightarrow r_0$
of the size of the trimer, and the corresponding redefinition
$Q_1 +g/3K \rightarrow Q_1$.  The breathing mode frequency $\omega_A$
in this model is $\sqrt{K/M}$;  in Na$_3$, multireference configuration
interaction calculations give this to be about 17 meV \cite{Busch},
while experiment \cite{Delacretaz} suggests 16 meV,
which implies $K\approx 1.5$ eV/\AA$^2$.

Unlike Eq. (\ref{eq:ha}), Eq. (\ref{eq:he})
cannot be simplified by a shift of origin. 
Rather than being a simple shifted parabola, the two-dimensional potential
surface of Eq. (\ref{eq:he}) has a cusp at the origin.  This is
the ``conical intersection'' between the lower energy adiabatic energy 
surface and the Jahn-Teller-split higher energy surface, which replaces
$-gQ$ by $+gQ$ in Eq. (\ref{eq:he}).  The higher surface
does not enter in adiabatic approximation.
The potential energy of Eq. (\ref{eq:he}) is cylindrically symmetric
around $Q_2=Q_3=Q=0$, and has the form of a ``Mexican hat.'' 
The value $Q_{\rm min}=2g/9K$ minimizes this potential
in the radial direction in $(Q_2,Q_3)$-space,
and the energy is lowered by the Jahn-Teller energy $-g^2/9K$.
The frequency of radial oscillation is $\omega_Q=\sqrt{3K/2M}$, but
for angular motion there is no restoring force,
$\omega_{\alpha}=0$.  The free motion
of the variable $\alpha$ is called ``pseudorotation.''  When
the coordinate $\alpha$ increases monotonically, the molecular coordinates
$r_n$ of the atoms move in counter-clockwise paths in the molecular $x,y$ 
plane (see Fig. \ref{fig:geom})
accompanied by no collective rotation, but by
true angular momentum, which of course must be quantized.

Quantum chemical calculations for Na$_3$ 
give a somewhat different picture, because 
the distortion ($\sim 10\%$) is not very small, and as a result
there is significant
warping of the path of least energy on the energy surface in the $Q_2,Q_3$ plane
surrounding the conical intersection at the origin.  Most calculations find the
minimum to be an obtuse isosceles triangle ($\alpha=180^{\circ}$
or equivalently $\alpha=\pm 60^{\circ}$ on Fig. \ref{fig:geom}).  
Saddle points, representing barriers for pseudorotation, occur 
at the acute positions ($\alpha=0^{\circ}$ or equivalently
$\pm 120^{\circ}$) \cite{Note2}.  The calculation of von Busch {\it et al.}
\cite{Busch} gives the average distortion $Q_{\rm min}=2g/9K\approx 0.25$ \AA.  
which implies that the coupling constant $g$ is about 1.7 eV/\AA.
The same calculation gives the energy minimum to be
97 meV and the barrier height to be 25 meV, higher than the
breathing phonon energy, and sufficient to strongly inhibit
pseudorotation.  In our model the energy minimum $-g^2/9K$ is about 200 meV,
twice the quantum-chemical value, and there is no barrier in leading order.

The kinetic part of Eq.(\ref{eq:he}) can be written as
$(3M/2)\dot{Q}^2+(I_Q/2)\dot{\alpha}^2$, in separated radial and
pseudorotational parts, with pseudorotational
moment of inertia $I_Q=3MQ^2$.
The zero-point amplitude of radial oscillation $(\hbar^2/6KM)^{1/4}$
is small enough compared with $Q_{\rm min}$ that we can approximate
the pseudorotational moment of inertia by the constant $I_Q=3MQ_{\rm
min}^2$.  Then the pseudorotational
wavefunction (in our approximation with no restoring force) is
$\psi_{\rm ps-rot}=\exp(im\alpha)$ and the energy is $\hbar^2m^2/2I_Q$.
As pointed out long ago by Longuet-Higgins {\it et al.} \cite{Longuet},
the pseudorotational quantum number $m$ must take half-integer values.
This is because the total wavefunction $\psi_{\rm ps-rot}(Q,\alpha)\psi_{\rm
ad}$ must be a single-valued
function of the coordinate $\alpha$.  We already noticed that the
adiabatic electron wave-function Eq. (\ref{eq:psiad})
changes sign under one pseudorotational
cycle.  Therefore the function $\psi_{\rm ps-rot}$ must also change
sign when $\alpha$ is increased by $2\pi$.  The ground state,
with pseudorotational quantum $m=\pm 1/2$,
is thus still doubly degenerate.  The electronic orbital degeneracy
was lifted by the Jahn-Teller distortion $Q \rightarrow Q_{\rm min}$, but
left and right pseudorotating states restore the
degeneracy of the combined electronic and atomic system.
This restoration of degeneracy is the discrete system analog of
restoration of translation invariance by phase solitons in one-dimensional
molecular chains \cite{Su}.

The wavefunctions $\psi_m$ of this adiabatic
vibrational ground state are products of the pseudorotational part
$\exp(im\alpha)$ times the electronic part $\psi_{\rm ad}$
(Eq.(\ref{eq:psiad}), also multiplied by the ground state
harmonic oscillator functions for breathing $Q_1$ and radial
oscillation $Q$.  The dipole operator in this 2d manifold is obtained by
projecting Eq. (\ref{eq:hefF}) onto $|m=\pm 1/2\rangle$
\begin{equation}
    {\cal H}_{F}^{\rm ad \ vib \ gs}=-\frac{\mu_0}{2}
    \vec{\tau}\cdot\vec{F}_m,
 \label{eq:mudotf1}
\end{equation}
where $\vec{\tau}=\tau_x \hat{x}+\tau_y \hat{y}$ 
are Pauli matrices acting in the space of
pseudorotational states $|m=1/2\rangle$ and $|m=-1/2\rangle$.
The eigenstates $\pm 1/2$ of pseudo-angular momentum ($-i\partial_{\alpha}$)
have left and right circulating dipolar charge currents.  Unlike the
undistorted molecule, the frequency of circulation is no longer $t_0/\hbar$,
but is now $\hbar/I_Q$, a much smaller number.  Therefore the orbital
magnetic dipole moment is restored, except with a much smaller value.
There is no diagonal expectation value of the electric dipole, only an
off-diagonal value, just as in the undistorted D$_{3h}$ case, but
reduced by 2, from $\mu_0$ to $\mu_0 /2$.  
Thus the half-filled outer orbital of the
symmetric triangular molecule has a quantum dipole which persists
under quantization of molecular distortions.
A field $\vec{F}_m$ applied in the plane
of the molecule splits the levels linearly, $\pm\mu_0|\vec{F}_m|/2$.
The eigenstates in the field are linear combinations of the
left and right pseudorotators whose moment is distributed in the
plane of the molecule with probability distribution $\cos^2(\alpha_F)$
where $\alpha_F$ is the angle measured from the projection of the
field onto the plane of the molecule.  This cosine distribution
is what causes the reduction of $\mu_0$ by 2.

In principle this is not fundamentally altered
by a warping potential like $(V_0/2)(1+\cos(3\alpha))$,
because tunneling along the  pseudorotational minimum energy
path preserves 3-fold symmetry and
guarantees a degenerate ground state \cite{Ham}.
Pseudorotational quantum numbers $m$ are still half-integer
quantized, but defined {\it modulo} 3, so that for example
$m=3/2$ and $m=-3/2$ are the same and this state is singly degenerate,
while $m=\pm 5/2$ is a doublet belonging to the $m=\mp 1/2$
representation.  The splitting $\hbar\omega_{\rm pseudorot}$
between the $m=\pm 1/2$ ground
state and the $m=3/2$ excited state 
gets reduced in frequency from $\omega=\hbar/I_Q$ 
by a Franck-Condon-type overlap factor governing the tunneling probability.
Thus, warping can greatly reduce the orbital magnetic dipole, 
which is proportional to the pseudorotational splitting.
Also, thermal mixing of pseudorotational states will smear
any effect for $k_B T > \hbar\omega_{\rm pseudorot}$.
%Careful experimental \cite{Busch}
%and theoretical \cite{Busch,Kendrick} studies confirm that the Na$_3$
%molecule evolves in two energetically equivalent
%modes of ``pseudorotation.''   

%%%%%%%%%%%%%%%%%%%%%%%%%%%%%%%%%%%%%%%%%%%%%%
%%%%%%%%%%%%%%%%%%%%%%%%%%%%%%%%%%%%%%%%%%%%%%

\section{Quantized rotations}\label{sec:rotations}

Now we must deal with real rotations of the molecule in space,
which adds a lot of complexity, as it does even for a
fixed dipole.  A full 3-d analysis does not fundamentally
change things from a much simpler
analysis of rotations in 2-d, so we confine ourselves to 2-d rotations in
this paper.  Here we
describe how 2-d rotations alter both a fixed dipole and our
quantum pseudorotator.  We work in the limit where the molecule stays
in its vibrational ground state, but has pseudorotational
excitations.  

The Hamiltonian for our pseudorotating quantum dipole (PQD) is
\bea
    {\cal H}_{\rm 2d \ PQD} &=&
    \frac{\hbar^2(\pi_{\alpha}+\pi_{\psi})^2}{2I_0}
    +\frac{\hbar^2 \pi_{\alpha}^2}{2I_Q} + \frac{V_0}{2}(1+\cos 3\alpha)
 \nonumber \\
    &-& \mu_0 F\cos(\psi+\alpha)
 \label{eq:2dqd}
\eea
Here $\psi$ is the Euler angle defining the orientation of the reference frame
of the molecule with respect to lab frame.  $\pi_{\alpha}$ and $\pi_{\psi}$
are dimensionless angular momenta which in quantum theory are
$-i\partial_{\alpha}$
and $-i\partial_{\psi}$ and take half-integer and integer values
correspondingly.
The first term of (\ref{eq:2dqd}) is the kinetic energy of rotation of the
trimer as a whole with $I_0 = 3Mr_0^2$.  The second term is the kinetic
energy of pseudorotation, and the third term is the simplest model
for the warped potential energy of pseudorotation, with minima at
the obtuse isosceles positions, and barriers of height $V_0$ at the
acute positions.  The simplest version of our model has
$V_0=0$.  The last term is the coupling to the
external electric field. It can be obtained from (\ref{eq:hefF}) by
substituting $\vec{F}_m=F(\cos\psi,-\sin\psi)$.

%The undistorted trimer is forbidden to have a classical dipole moment but has a
%quantum one as we have shown in Sec. \ref{sec:trimer}. If the molecule is
%distorted, say, by Jahn-Teller effect it may acquire a classical dipole
%moment. Obviously, there must be a crossover regime between quantum and
%classical behavior of the electric dipole moment governed by the strength
%of the Jahn-Teller distortion.
                                                                                          
%In this section we study the effects of the ``quantum nature'' of the dipole
%moment. For this purpose we consider the limit of a very small distortion
%($Q_{\rm min}\ll r_0$).  

%We study the effects of spatial rotations and finite
%temperature on the observable electric dipole moment of the trimer in this
%limit. We take model 

Eq. (\ref{eq:2dqd}), with $V_0$ set to zero,
models a quantum electric dipole moment, free to rotate in 2d space.
We now calculate the observable dipole moment for a statistical ensemble
at temperature $T$.
This is $\langle\mu\rangle =-\partial {\cal F}/\partial F $, where
${\cal F}$ is the free energy of the dipole calculated at temperature $T$ with
Hamiltonian (\ref{eq:2dqd}).  The temperature scales set by
rotations and pseudo-rotations are
$T_{\rm rot}=\hbar^2/2I_0\ll T_{\rm ps-rot}=\hbar^2/2I_Q$.
For Na$_3$ these are approximately 0.1K and 5K.
We assume $T\gg T_{\rm rot}$, 
which allows us to treat spatial rotations classically,
but $T$ can be either above or below $T_{\rm ps-rot}$.
                                                                                          
Before proceeding to the quantum dipole (\ref{eq:2dqd}) we first summarize the
well known results for the classical dipole moment in an external electric field.

%%%%%%%%%%%%%%%%%%%%%%%%%%%%%%%%%%%%%%%%%%%%%%

%
\begin{figure}[t]
   \includegraphics[height=3.0in,width=3.0in,angle=0]{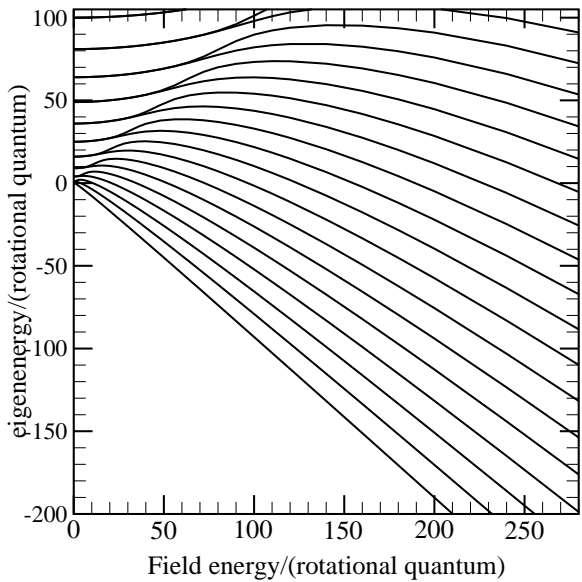}
   \caption{Spectrum of low-lying energy eigenvalues of the 2-d
fixed dipole rotator.  The field energy is $\mu_0 F$ and the
rotational quantum of energy is $T_{\rm rot}=\hbar^2 /2I$.
The zero field axis has dimensionless eigenvalues equal to $m^2$
for $m=0,\pm 1,\pm 2,\ldots$.  Each level except $m=0$ is doubly degenerate,
and the degeneracy is split by the field.}
\label{fig:2dfd}
\end{figure}

\subsection{Classical dipole} \label{sec:cd}

The Hamiltonian of a 2d fixed dipole is
\begin{equation}
    {\cal H}_{\rm 2d \ FD} = \frac{\hbar^2 \pi_{\psi}^2}{2I_0} -\mu_0 F
\cos\psi
 \label{eq:2dfd}
\end{equation}
where $I_0$ is the molecule's moment of inertia.  This
is the pendulum problem.  The low-lying quantum eigenenergies of this
system, calculated numerically, are shown in Fig. \ref{fig:2dfd}.

The classical treatment is standard.  After evaluating
the partition function $Z=\exp(-{\cal F}/T)$ (we use units $k_B=1$), the
classical average dipole is
found by differentiating ${\cal F}$ to be
\begin{equation}
    \langle\mu\rangle=\mu_0 \frac{\int_{-\pi}^{\pi}d\psi \cos\psi e^{(\mu_0
F/T)\cos\psi}}
    {\int_{-\pi}^{\pi}d\psi e^{(\mu_0 F/T)\cos\psi}}
 \label{eq:zcl}
\end{equation}
which saturates to $\mu_0$ at
high field $F$ and equals $\chi F$ at low field, where $\chi=\mu_0^2 /2 T$
is the 2-d version of the Debye-Langevin polarizability \cite{Vanvleck}.

\begin{figure}[t]
   \includegraphics[height=3.0in,width=3.0in,angle=0]{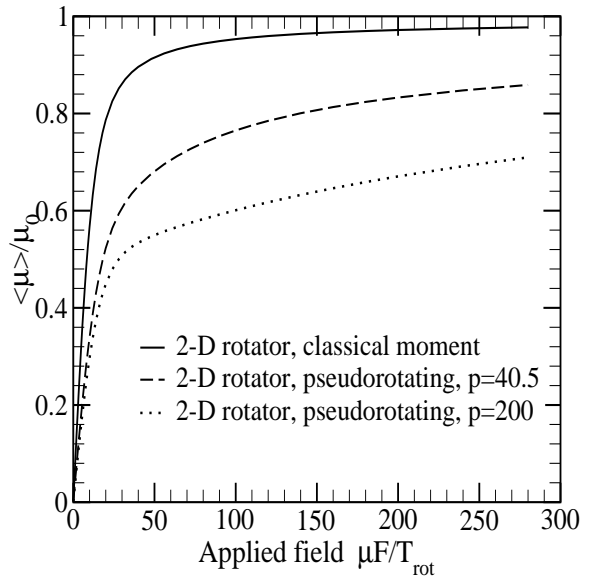}
\caption{Dipole moment of classical and quantum dipoles as a
function of applied electric field, calculated at temperature
$T=7.0T_{\rm rot}$.  The quantum rotator has a dimensionless
parameter $p=T_Q/T_{\rm rot}$ which is set to 40.5 (dashed curve,
reasonable for Na$_3$) and to 200 (dotted curve).}
\label{fig:mu}
\end{figure}

The results of the quantum treatment
of this problem do not have a simple form.  At high field
($\mu_0 F \gg  T$) the dipole makes pendulum oscillations around
the direction of the field.  This is seen in the lower right part
of Fig. \ref{fig:2dfd} where there are evenly spaced levels
at $\Delta E=\hbar\omega$ where the pendulum frequency $\omega=\sqrt{\mu_0
F/I_0}$ increases as the field increases.  
At low field and low temperature, the quantum ground state
is rotationally symmetric with a dipole induced by second-order coupling
to the first rotationally excited states.  In this limit the
polarizability becomes $\mu_0^2/(\hbar^2/2I)$.  If the field increases
to values large compared with $T/\mu_0$, the eigenstates are mixtures
of rotational levels.  These mixtures give a full dipole moment, saturated at the
value $\langle\mu\rangle = \mu_0$.  We plot the dependence of dipole moment
versus an applied electric field in Fig. \ref{fig:mu}. In different limits
we have
\begin{eqnarray}
    \frac{\langle\mu\rangle_{\rm 2d \ CD}}{\mu_0} &\approx&
\frac{\mu_0F}{T_{\rm rot}}
    \;\;\; (\mu_0 F \ll T_{\rm rot})
 \nonumber \\
     &\approx&  \frac{\mu_0F}{2T}
    \;\;\; (T_{\rm rot}\ll \mu_0 F \ll T)
 \nonumber \\
    &\approx& 1-\frac{ T}{\mu_0 F}
    \;\;\; (T\ll  \mu_0 F \ll \frac{T^2}{T_{\rm rot}})
 \label{eq:cldip} \\
    &\approx& 1-\frac{1}{4}
    \sqrt{\frac{2T_{\rm rot}}{\mu_0 F}}
    \;\;\; (\mu_0 F \gg \frac{T^2}{T_{\rm rot}}),
 \nonumber
\end{eqnarray}
where the subscript ``CD'' means classical dipole (the rotations
are treated quantum mechanically).  The water molecule
is a good example.  Normally we think of a classical fixed dipole,
but the quantum eigenstates in zero field are rotating dipoles with no
fixed average moment.  In a typical chemical or biological environment,
the environmental field $F$ is large, free rotations are hindered,
and the dipole is exposed.

%

%%%%%%%%%%%%%%%%%%%%%%%%%%%%%%%%%%%%%%%%%%%%%%
\subsection{Quantum dipole with free pseudorotations}\label{sec:qdf}
                                                                                          
To emphasize the difference between a quantum dipole (\ref{eq:2dqd}) and
a fixed dipole (\ref{eq:2dfd}), we choose the limit
$V_0=0$, i.e., a quantum dipole with free pseudorotations.
Eq.(\ref{eq:2dqd}) with $V_0=0$ is a funny degenerate sort of double
pendulum
with two angular degrees of freedom.
The quantum states at zero field $\vec{F}=0$ have eigenvalues
\begin{equation}
    \epsilon(\ell,m)=\hbar^2 (\ell+m)^2/2I_0 +\hbar^2 m^2/2I_Q
 \label{eq:zfeig}
\end{equation}
where the angular momentum quantum number $\ell$ takes integer
values $0,\pm1,\ldots$ and the pseudomomentum quantum number
$m$ takes half-integer values $\pm 1/2, \pm 3/2, \ldots$.  Because
$I_Q \ll I_0$, the low-lying states have $m=\pm 1/2$.  All
eigenstates are 4-fold degenerate, even the ground state, because
for either choice of $|m|$ there are two ways to get any allowed
value of $|\ell + m|$.  When the field is applied, each quadruplet
level splits into two doublets.  There are no allowed matrix
elements of the field in any degenerate manifold of 4 states, so
the initial splitting is quadratic in field, just as for water.
If we assume that the temperature is low enough that only the lowest
pseudorotational levels $m=\pm 1/2$ are excited, it is a good approximation
to ignore the term in Eq.(\ref{eq:2dqd}) involving $I_Q$.
Then the 2-d Hamiltonian (\ref{eq:2dqd})
separates into $2\times 2$ matrices acting on subspaces formed
by $|l,m=1/2\rangle$, $|l-1,m=-1/2\rangle$.   The ground state energy
is $T_{\rm rot}[5/4-\sqrt{1+(\mu_0 F /(2 T_{\rm rot}))^2}]$, which implies
a ground state dipole $\langle\mu\rangle=(\mu_0 /2)[(\mu_0 F/2)/\sqrt{(\mu_0
F/2)^2
+T_{\rm rot}^2}]$.  For high fields and $T$ not too low
($\mu_0 F \gg T \gg T_{\rm rot}$) the dipole is
\begin{equation}
    \langle\mu\rangle_{\rm 2d \ QD} \approx \frac{\mu_0}{2} \left[
    1-\frac{2 TT_{\rm rot}}{(\mu_0 F/2)^2} \right]
 \label{eq:qudip}
\end{equation}
\begin{figure}[t]
   \includegraphics[height=3.0in,width=3.0in,angle=0]{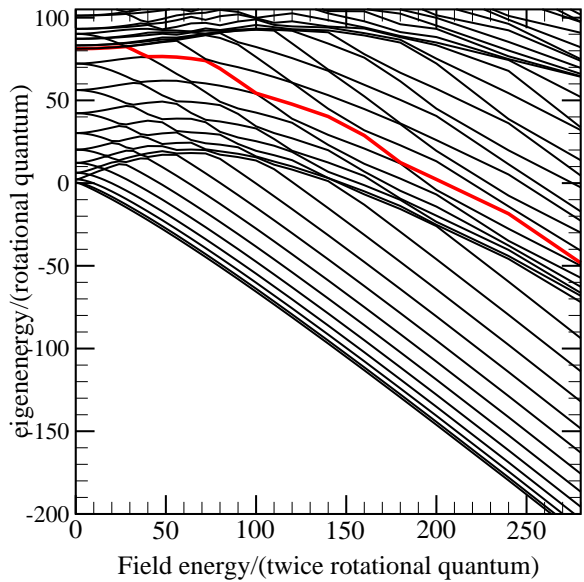}
   \caption{Spectrum of low-lying energy eigenvalues of the 2-d
quantum dipole rotator.  The field energy is $\mu_0 F$ and the
rotational quantum of energy is $\hbar^2 /6Mr_0^2$.
The zero field axis has dimensionless eigenvalues equal to $(\ell+m)^2
+(r_0/q_{\rm min})^2 m^2$, with the zero-point pseudorotational
quantum $(r_0/q_{\rm min})^2 (1/2)^2$ subtracted out.
The level in red at zero $F$ is the first with pseudorotational
quantum number $|m|>1/2$.
Each level at zero $F$ is quadruply degenerate,
and the degeneracy is reduced to double by the field.}
\label{fig:2dqd}
\end{figure}

Notice that the effect of temperature (when it is less than
the field energy $\mu_0 F$) is smaller on the quantum (pseudo- + real-)
rotating dipole Eq.(\ref{eq:qudip}) than on the fixed rotating
dipole Eq.(\ref{eq:cldip}); there is an extra small factor
$T_{\rm rot}/(\mu_0 F/2)$ which postpones thermal degradation
of the dipole by rotation.  The reason is evident in the difference
between the spectra seen in Fig. \ref{fig:2dfd} and Fig. \ref{fig:2dqd}.
At higher fields $F$ and not too high excitation energy,
the fixed dipole has energy levels that fan out
to larger separations in higher fields, meaning lower slopes and lower
dipoles in higher excited states,
while the quantum dipole has energy levels whose slopes
do not depend on excitation level.  The fixed dipole levels
correspond to pendulum oscillations of the dipole away from the
field direction, whereas the quantum dipole levels correspond to
free rotations (the energy levels are $\hbar^2(\ell+1/2)^2 /2I_0 $)
of a molecule whose dipole is fixed in the direction of the field.
The way this happens is that the dipole is not fixed in the axes
of the molecule, but free to pseudorotate around these axes.  Real
rotations and pseudorotations give compensating changes in the dipole
moment, leaving the moment fixed in space.  This happens equally well
for left and right rotations, so all levels are still doublets.
The comparison between dependences of $\langle\mu\rangle$ vs. $F$ is
presented in Fig. \ref{fig:mu}.
                                                                                          
%%%%%%%%%%%%%%%%%%%%%%%%%%%%%%%%%%%%%%%%%%%%%%
%%%%%%%%%%%%%%%%%%%%%%%%%%%%%%%%%%%%%%%%%%%%%%

\section{Discussion and conclusions}
\label{sec:concl}

The main point of the present paper is the 
possible relevance to the issue of
unusual polarity in metal clusters \cite{Moro1,Moro2}.
Existing experiments on X$_3$ molecules have not seen the
effects described here.  There are multiple possible reasons,
including (a) not looking for quite the right effect,
(b) need for lower temperature, (c) nuclear exchange
symmetry, and (d) suppression by large barriers to pseudorotation.
A subsequent paper will address some of these issues in the context
of Na$_3$.  Here we speculate about the more general issues.

Suppose we discard triangular symmetry
altogether, as in a mixed alkali trimer like
Li$_2$Na \cite{Deshpande}, or LiNaK.
In the symmetric case, symmetry forced the conical intersection
which preserved (by pseudorotation) the degenerate ground state
and its quantum dipole.  However, as explained by Wigner
\cite{Wigner,Herzberg}, points
of degeneracy, or conical intersections, are generically present
\cite{Yarkony} if there are enough free variables.  In the last occupied
molecular orbital level of LiNaK, the splitting of the two
sublevels can be tuned to zero by adjustment of two parameters,
since the Hamiltonian matrix can be written in this manifold
as real symmetric $2\times2$.  There are three such
parameters, the normal mode amplitudes $(Q_1,Q_2,Q_3)$.  Thus
it is permitted to have a line in configuration space where the upper
and lower energy surfaces touch and away from which they
separate linearly.  Consider the minimum energy point on this line,
and also the global minimum of the lower energy surface.
Away from the global minimum, the energy increases quadratically
with three principal axes defining nearby ellipsoidal
energy surfaces.  Now begin moving along the axis of least
energy increase, and continue following the bottom of this valley.  
It is possible that this valley will return to the global
equilibrium after circling the line of accidental degeneracy.
It is also possible that the path in this minimum valley
lies always lower than the minimum energy found on the line of
degeneracy.  In this case, there is again a pseudorotation
with an electronic-wavefunction sign change.  The lowest
energy pseudorotational state must therefore have a compensating
phase change of $\pi$ when orbiting the
degenerate line.  Time-reversal symmetry then requires that
an inequivalent reverse pseudorotation with opposite phase
is degenerate.  These two states should each drag a dipole
moment in opposite directions once around the degeneracy line
as the system moves adiabatically around the circuit.  Thus
a quantum dipole is generically permitted.  Larger clusters
of metal atoms have more complex energy surfaces, and both
features (1) conical intersections and (2) lack of inversion
symmetry are generically likely.  Therefore unusual dipoles
whose properties are not fully classical may be expected.
Possibly this is what Moro {\it et al.} \cite{Moro1,Moro2} observed.
                                                                                          
Moro {\it et al.} observed a correlation between unusual
dipoles and metals known in bulk form to have
superconductivity caused by large electron-phonon coupling.
Large electron-phonon coupling derives from a high electronic
density of states and a rapid change of single-particle
energy eigenvalues with deformation.  These are conditions
favorable to the occurence of the dipole we have described.
A high density of states means plentiful levels whose energies
may accidentally cross at conical intersections.  Sensitivity
of eigenvalues to distortion means that the point of the cone
of intersection can lie higher in energy than the surrounding
landscape.  Thus our picture contains the right ingredients.
Further experiment and more accurate microscopic modelling
will be needed to confirm it.

To conclude, we studied the formation of an electrical dipole moment in
triangular molecules. Although, the classical dipole moment is forbidden for
symmetric trimers it is formed due to the quantum degeneracy of the ground
states.  This dipole moment survives Jahn-Teller distortion of the molecule
and keeps its ``quantum nature'' for small distortions. 
It is interesting to note that the quantum dipole moment makes a
two-level system which can be manipulated by an external electric field. If
it were possible to have a controlled interaction between such trimers, one
could investigate the use of this two-level system for quantum computing
purposes similarly to what was done for quantum qubits using persistent
currents\cite{Kulik}.

%%%%%%%%%%%%%%%%%%%%%%%%%%%%%%%%%%%%%%%%%%%%%%
%%%%%%%%%%%%%%%%%%%%%%%%%%%%%%%%%%%%%%%%%%%%%%

\section{Acknowledgements}

We thank W. de Heer for discussions of ref. 1.  
We acknowledge helpful suggestions from T. Baruah, W. Ernst,
K. Lehmann, J. Muckerman, R. Porter, G. Scoles, and A. M. Stoneham.  The work of PBA
was supported in part by NSF grant no. DMR-0089492, and in part
by a J. S. Guggenheim Foundation fellowship.  Work at
Columbia was supported in part by the MRSEC Program of the NSF
under award no. DMR-0213574.  Work of AGA was supported by
the Alfred P. Sloan foundation, NSF grant DMR-0348358, and the Theory Institute
of Strongly Correlated and Complex Systems at Brookhaven.

%%%%%%%%%%%%%%%%%%%%%%%%%%%%%%%%%%%%%%%%%%%%%
%%%%%%%%%%%%%%%%%%%%%%%%%%%%%%%%%%%%%%%%%%%%%

\appendix
\section{\label{app:bloch}Bloch sphere}
It is convenient to use the Bloch sphere parameterization of the degenerate
ground state $|g\rangle=(\alpha,\beta)^{t}
=\alpha|k=-1\rangle +\beta|k=1\rangle$.  Define the unit vector
$\vec{n}=\langle g|\vec\sigma|g\rangle$. The manifold of degenerate states
of the two-level system (undistorted trimer) is represented then by
two-dimensional sphere swept by the unit vector $\vec n$. The electric
dipole moment $\vec\mu$ in the corresponding ground state is proportional to
the projection $\vec{n}_{\perp}$ of $\vec n$ on the $xy$ plane $\vec\mu
=\mu_{0}\vec{n}_{\perp}$.

\begin{figure}[t]
   \includegraphics[height=2.1in,width=2.1in,angle=0]{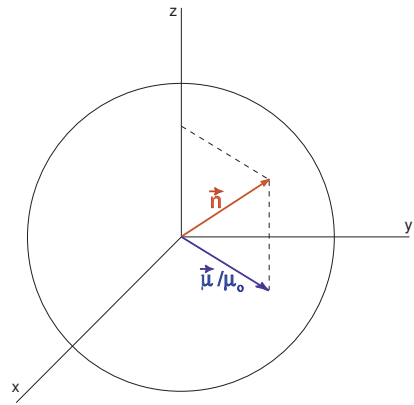}
   \caption{\label{fig:bloch} In the Bloch sphere parameterization the
ground states of two-level system correspond to the points of a unit sphere
i.e., they can be labeled by a unit vector $\vec n$. For the undistorted
trimer the $xy$ projection of this unit vector is proportional to the
electric dipole moment while its $z$ component is proportional to the
magnetic moment of the ground state.}
\end{figure}
%

%xxxxxxx

The partly-occupied states
$|k=\pm 1\rangle$ of the static trimer 
have an orbital magnetic moment perpendicular to the plane of the
molecule which couples to a magnetic field $\vec{B}$.
In the degenerate $E^{\prime}$ subspace, the coupling is
\begin{equation}
    {\cal H}_{B}= -\mu_{\rm orb} \sigma_z B_z,
 \label{eq:hmag}
\end{equation}
where the moment $\mu_{\rm orb}$ is $2\pi e\omega A$ as previously discussed.
In the Bloch sphere parameterization, the magnetic dipole moment is
proportional to the projection of $\vec{n}$ onto the $z$ axis, and is equal to
$\mu_{\rm orb}n_{z}$.

The value of the moment $\mu_{\rm orb}$ for Na$_3$ is about 0.3 Bohr magnetons,
if we ignore the barrier to free pseudorotation.
The electrical moment $er_0 /2$ is 0.98 e\AA \ or 4.7D (1 Debye unit =
0.208 e\AA.) in our naive model.  An unpublished density functional
calculation \cite{Tunna} gives 2.3D.  It
is not surprising that the actual result is reduced,
because the true molecular orbitals surely have a significant $p\sigma$
component missing from our model.
Also, the on-site Coulomb repulsion causes a suppression
of the dipole as is shown in the next appendix.

%%%%%%%%%%%%%%%%%%%%%%%%%%%%%%%%%%%%%%%%%%%%%
%%%%%%%%%%%%%%%%%%%%%%%%%%%%%%%%%%%%%%%%%%%%%

\section{\label{app:hubbard}Effect of a Hubbard Term}

\begin{figure}[htb]
\begin{center}
\includegraphics[width=0.4\textwidth]{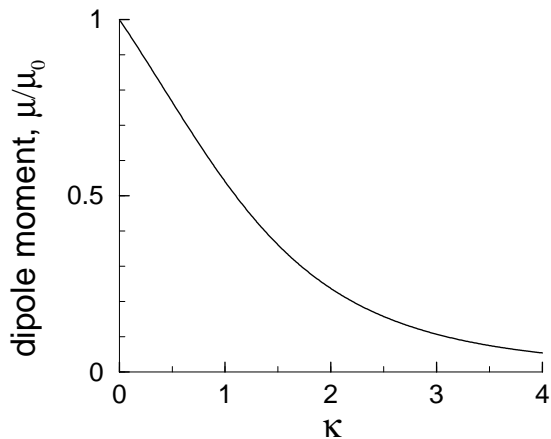}
\caption{The quantum dipole in units $e r_0/2$
as a function of the dimensionless Hubbard strength
$\kappa \equiv U/t_0$. The intercept
coincides with the non-interacting H\"{u}ckel
model, while for $\kappa \gg 1$ it
falls as $\kappa^{-3}$. }
\label{fig:hubbard}
\end{center}
\end{figure}
The Coulomb interaction between electrons is approximated by a
Hubbard term of strength $U$,
\ba
H_{int} = U \, \sum_{n = 0,1,2} \;
\LB c^{\dag}_{n\ua} c_{n\ua} + c^{\dag}_{n\da} c_{n\da} \RB^2
\label{eq:hubbard}
\ea
The Hilbert space has dimension 20. There are 8 basis states
with no vacant sites and 12 with one vacant site. Total spin, spin
$\hat{z}$-projection, and orbital (electronic) angular momentum
are conserved. In the eight-dimensional $S=1/2$ and
$S_z=1/2$ subspace, there are three states with
angular momentum $k=1$, three
with $k=-1$, and two with $k=0$ that have higher
energy. Parity symmetry is not broken by the Hubbard term.
The ground state is a $k=\pm 1$ doublet, as it was for the
non-interacting model.
Figure \ref{fig:hubbard} shows the dependence of the quantum dipole
on the dimensionless Hubbard parameter
$\kappa \equiv U/t_0$.  A factor of two suppression occurs when
$U$ is of the same order as $t_0$.
For $\kappa \ll 1$, the doubly degenerate ground state
energy is $E_0 \approx -3 t_0 + (13/3) U$.  The dipole operator
has an off-diagonal matrix element $\langle -1|\vec{\mu}|+1\rangle$
between $k=-1$ and $k=+1$ states equal to $-\mu (\hat{x}-i\hat{y})$
with
\begin{equation}
\mu \approx\mu_{0}\left[1 -\frac{4}{9}\kappa \right], \;\;\;\mbox{for}\;
\kappa\ll 1.
\end{equation}
This agrees with the non-interacting result, Eq. (\ref{eq:mu}).
For $\kappa \gg 1$, the doubly degenerate ground state has
energy $E_0 \approx 3 U - 3 t_{0}^2/U$.  The magnitude $\mu$ of
the off-diagonal matrix element falls rapidly as shown in Fig. \ref{fig:hubbard},
going asymptotically as
\ba
\mu \approx  \frac{9}{2} \kappa^{-3} \;\mu_0, \;\;\;\mbox{for}\; \kappa\gg 1.
\ea
%

%%%%%%%%%%%%%%%%%%%%%%%%%%%%%%%%%%%%%%%%%%%%%%
%%%%%%%%%%%%%%%%%%%%%%%%%%%%%%%%%%%%%%%%%%%%%%

%%%%%%%%%%%%%%%%%%%%%%%%%%%%%%%%%%%%%%%%%%%%%
%%%%%%%%%%%%%%%%%%%%%%%%%%%%%%%%%%%%%%%%%%%%%

\begin{thebibliography}{99}
\bibitem{xxx} [Requist, Baruah, Abanov, Allen] -- lead
author to be decided later.
\end{thebibliography}
\end{document}